\documentclass[11pt]{article}
\usepackage{axodraw}
\usepackage{epsfig}
\usepackage{amsfonts}
\usepackage{amsmath}
\allowdisplaybreaks[1]

\input pix.sty

\newcommand{\Xs}{{X}^{ }_1}
\newcommand{\Xo}{{X}^{ }_8}
\newcommand{\Ss}{{S}^{ }_1}
\newcommand{\So}{{S}^{ }_8}
\newcommand{\bSs}{\bar{S}^{ }_\rmi{1}}
\newcommand{\bSo}{\bar{S}^{ }_\rmi{8}}

\renewcommand{\eq}{eq.~}

\renewcommand{\se}{sec.~}

\renewcommand{\fig}{fig.~}

\newcommand{\alphas}{\alpha_{\rm s}}
\newcommand{\Nf}{N_{\rm f}}
\newcommand{\Nc}{N_{\rm c}}

\newcommand{\CF}{C_\rmii{F}}

\def\lsi{\raise0.3ex\hbox{$<$\kern-0.75em\raise-1.1ex\hbox{$\sim$}}}
\def\gsi{\raise0.3ex\hbox{$>$\kern-0.75em\raise-1.1ex\hbox{$\sim$}}}
\newcommand{\lsim}{\mathop{\lsi}}

\newcommand{\nF}{n_\rmii{F}}
\newcommand{\nB}{n_\rmii{B}}
 \renewcommand{\nF}[1]{f_\rmii{F{#1}}}
 \renewcommand{\nB}[1]{f_\rmii{B{#1}}}

\newcommand{\rmii}[1]{{\mbox{\tiny\rm{#1}}}}

\newcommand{\re}{\mathop{\mbox{Re}}}
\newcommand{\im}{\mathop{\mbox{Im}}}

\newcommand{\Tint}[1]{{\hbox{$\sum$}\!\!\!\!\!\!\!\int\,}_{\!\!\!\!\raise-0.9ex\hbox{$\scriptstyle{#1}$}}}
\newcommand{\Tinti}[1]{{{\Sigma}\!\!\!\!\raise0.3ex\hbox{$\int$}_\rmii{${#1}$}}}

\newcommand{\bi}{\begin{itemize}}
\newcommand{\ei}{\end{itemize}}
\newcommand{\hide}[1]{ }

\newcommand{\ff}{\rmi{\sl f\,}}

\newcommand{\deltabar}{\delta\!\!\!\raise0.7ex\hbox{--}\,}
\def\TAsc(#1,#2)(#3,#4,#5)%
{\SetWidth{2.0}\CArc(#1,#2)(#3,#4,#5)\SetWidth{1.0}}
\def\Lwidth{3}

\def\TAgl(#1,#2)(#3,#4,#5){\SetWidth{2.0}\PhotonArc(#1,#2)(#3,#4,#5){\Lwidth}%
{6.283 #3 mul 360 div #4 #5 sub #4 #5 sub mul sqrt mul Tdensity mul}%
\SetWidth{1.0}}
\def\TLgl(#1,#2)(#3,#4){\SetWidth{2.0}\Photon(#1,#2)(#3,#4){\Lwidth}
{#1 #3 sub #1 #3 sub mul #2 #4 sub #2 #4 sub mul add sqrt Tdensity mul}%
\SetWidth{1.0}}
\newcommand{\piC}[1]{\;\parbox[c]{120pt}{\begin{picture}(120,60)(0,0)
\SetWidth{1.0}\SetScale{1.2} #1 \end{picture}}\; }
\renewcommand{\picc}[1]{\;\parbox[c]{60pt}{\begin{picture}(60,30)(0,0)
\SetWidth{1.0}\SetScale{1.3} #1 \end{picture}}\; }
\def\Lwidth{1.3}

\def\NRa{\piC{%
 \SetWidth{1.5} 
 \Line(30,0)(50,0)%
 \Line(30,5)(50,5)%
 \Line(70,0)(90,0)%
 \Line(70,5)(90,5)%
 \Line(30,30)(90,30)%
 \Line(30,25)(90,25)%
 \CArc(90,15)(10,-90,90)%
 \CArc(90,15)(15,-90,90)%
 \CArc(30,15)(10,90,270)%
 \CArc(30,15)(15,90,270)%
 \DashLine(52,2.5)(68,2.5){1}%
 \COval(51,2.5)(7,3)(0){Black}{Yellow}%
 \COval(69,2.5)(7,3)(0){Black}{Yellow}%
 \CBoxc(14.5,15)(5,5){Black}{Blue}%
 \CBoxc(105.5,15)(5,5){Black}{Blue}%
}}
\def\NRb{\piC{%
 \SetWidth{1.5} 
 \Line(30,0)(50,0)%
 \Line(30,5)(50,5)%
 \Line(70,0)(90,0)%
 \Line(70,5)(90,5)%
 \Line(30,30)(90,30)%
 \Line(30,25)(90,25)%
 \CArc(90,15)(10,-90,90)%
 \CArc(90,15)(15,-90,90)%
 \CArc(30,15)(10,90,270)%
 \CArc(30,15)(15,90,270)%
 \DashLine(52,2.5)(68,2.5){1}%
 \COval(51,2.5)(7,3)(0){Black}{Yellow}%
 \COval(69,2.5)(7,3)(0){Black}{Yellow}%
 \CBoxc(14.5,15)(5,5){Black}{Blue}%
 \CBoxc(99.0,15)(5,5){Black}{Blue}%
}}
\makeatletter \@addtoreset{equation}{section} \makeatother

\makeatletter
\renewcommand\section{\@startsection {section}{1}{\z@}%
                                   {-5.5ex \@plus -1ex \@minus -.2ex}% bfr-
                                   {2.3ex \@plus.2ex}%
                                   {\normalfont\large\bfseries}}
\renewcommand\subsection{\@startsection{subsection}{2}{\z@}%
                                     {-3.25ex\@plus -1ex \@minus -.2ex}%
                                     {1.5ex \@plus .2ex}%
                                     {\normalfont\normalsize\bfseries}}
\renewcommand\thesection {\@arabic\c@section}
\renewcommand\thesubsection   {\thesection.\@arabic\c@subsection}
\renewcommand{\@seccntformat}[1]{%
\csname the#1\endcsname.\hspace{1.0em}}
\makeatother
\begin{document}

\flushbottom

\begin{titlepage}

\begin{flushright}
% DRAFT \\ 
BI-TP 2012/43\\
% arXiv:yymm.nnnn\\ 
\vspace*{1cm}
\end{flushright} 
\begin{centering}
\vfill

{\Large{\bf
Sommerfeld effect in heavy quark chemical equilibration
}} 

\vspace{0.8cm}

D.~B\"odeker$^{\rm a}$, %\footnote{bodeker@physik.uni-bielefeld.de}, 
M.~Laine$^{\rm b}$ %\footnote{laine@itp.unibe.ch}

\vspace{0.8cm}

$^\rmi{a}$%
{\em
Faculty of Physics, University of Bielefeld, 
D-33501 Bielefeld, Germany\\}

\vspace*{0.3cm}

$^\rmi{b}$%
{\em
Institute for Theoretical Physics, 
Albert Einstein Center, University of Bern, \\ 
  %for Fundamental Physics, 
Sidlerstrasse 5, CH-3012 Bern, Switzerland\\}

\vspace*{0.8cm}

\mbox{\bf Abstract}
 
\end{centering}

\vspace*{0.3cm}
 
\noindent
The chemical equilibration of heavy quarks in a quark-gluon plasma proceeds
via annihilation or pair creation.  For temperatures $ T $ much below the
heavy quark mass $ M $, when kinetically equilibrated heavy quarks move very
slowly, the annihilation in the colour singlet channel is enhanced because the
quark and antiquark attract each other which increases their probability to
meet, whereas the octet contribution is suppressed. This is the so-called
Sommerfeld effect. It has not been taken into account in previous calculations
of the chemical equilibration rate, which are therefore incomplete for $ T
\lsim \alphas^2 M $. We compute the leading-order equilibration rate in this
regime; there is a large enhancement in the singlet channel, but the rate is
dominated by the octet channel, and therefore the total effect is small. In
the course of the computation we demonstrate how operators that represent the
annihilation of heavy quarks in non-relativistic QCD can be incorporated into
the imaginary-time formalism.

\vfill

%\noindent
%PACS numbers: 
%11.10.Wx, %        Finite temperature field theory
%11.15.Ha, %        Lattice gauge theory 
%12.38.Bx, %        Perturbative calculations in QCD
%12.38.Mh, %        Quark--gluon plasma
%14.40.Nd, %        Bottom mesons
%\\
%Keywords:
 
\vspace*{1cm}
  
\noindent
October 2012

\vfill

\end{titlepage}

%%%%%%%%%%%%%%%%%%%%%%%%%%% SECTION %%%%%%%%%%%%%%%%%%%%%%%%%%%%%%%%%%%%%%
%
\section{Introduction}

If heavy quarks of mass $ M $ in a quark-gluon plasma are initially
out of thermal equilibrium, they quickly equilibrate kinetically by
multiple scatterings with gluons and light 
quarks~\cite{mt}--\cite{mumbai}. 
(In this paper we consider an ideal limit in which the plasma
lives for a long time.) At temperatures 
$ T \ll M $ chemical equilibration is much slower because it 
requires quark-antiquark annihilation or pair creation.%
 \footnote{Weak interactions are not considered here.}  
In fact, given that the probability to find a target is Boltzmann suppressed, 
the chemical equilibration rate $ \Gamma_\rmi{chem} $  
is exponentially small at low temperatures, 
$ \Gamma_\rmi{chem} \sim e ^{ - M/T }$~\cite{s1,s2,msmc}.

Given that the heavy quarks are in kinetic equilibrium they move with
non-relativistic velocity. When the
reacting particles have small relative velocity $ v $, their mutual
interactions can have a large influence on the annihilation or 
production cross section~\cite{sommerfeld,landau3}, a phenomenon
known as the Sommerfeld effect. In
perturbation theory this effect would first show up in the 1-loop
correction to a tree-level cross section $ \sigma  _ 0 $:
\be
 \sigma = \sigma_0 \, 
 \Bigl[ 
   1 + O\Bigl( \frac{\alphas}{v} \Bigr)
     + O\bigl( {\alphas} \ln {v} \bigr)
     + O\bigl(\alphas\bigr)
 \Bigr]
 \;. \la{sketch}
\ee
For Coulomb-like interactions the Sommerfeld effect manifests itself
as a non-vanishing  contribution of  
$O(\alpha_s/v)$.  When $ v $ becomes as small
as $ \alphas $,  the 1-loop correction can become larger than the tree-level
result. The naive loop expansion then breaks down, and the
enhanced terms need to be resummed. 
This effect has to be taken into account for any particle reactions 
close to threshold~\cite{fadin-top}, and has been widely discussed, e.g.,\ 
in connection with $t\bar{t}$ and gluino and squark pair production 
in hadronic collisions (see e.g.\ refs.~\cite{%
  Beneke:2012gn,Langenfeld:2012ti,Beenakker:2011gt} 
for recent work and references). 
It may also play an important role in the indirect detection 
of dark matter particles~\cite{hisano-explosive}.

At finite temperature the typical heavy quark velocity 
is of order $ v \sim \sqrt{ T/M } \ll 1$. Thus the naive
perturbative expansion breaks down for 
$v \lsim \alphas$, i.e.\ $ T \lsim \alphas ^ 2 M $. 
Then a similar resummation  is needed 
as in hadronic collisions at zero temperature. 
The Sommerfeld effect in thermal 
dark matter freeze-out has indeed been discussed 
in many recent works, such as refs.~\cite{%
  hisano-thermal,feng,Hryczuk:2010zi}, and 
it may also play a role in certain leptogenesis 
scenarios~\cite{Strumia:2008cf}. 
However, to the best of our knowledge, the Sommerfeld effect
has not been taken into account in previous calculations of the
chemical equilibration rate of heavy quarks.

In ref.~\cite{chemical}, a definition of the heavy quark 
chemical equilibration rate 
was given which is non-perturbative and thus goes beyond the usual
formulation in terms of the Boltzmann equation. At leading order it
gives the same rate as the Boltzmann equation. To include more terms
of the perturbative expansion, it would be convenient to use
non-relativistic QCD (NRQCD~\cite{nrqcd}) for computing the rate. The pair
annihilation of heavy quarks is represented in NRQCD by an imaginary
part in a coefficient of a 4-fermion operator~\cite{bodwin}. Such a complex
coefficient is %, obviously, not 
related 
%to CP-violation, but 
to the analytic 
structure of a corresponding Green's function. 
% In the imaginary-time formalism no imaginary part appears. 
One purpose of the present paper is to give a formulation of 
such operators which can be used in the imaginary-time formalism. 
Subsequently, 
the NRQCD analysis allows us to disentangle the contributions from
the colour singlet and octet operators to the heavy quark chemical 
equilibration rate, a necessary first step for discussing the 
Sommerfeld effect. 

This paper is organized as follows. 
In~\se\ref{sc:nrqcd} we discuss how 
pair annihilation can be incorporated in the imaginary-time formulation
of NRQCD, and use this to determine the contributions of singlet and 
octet operators to the chemical equilibration rate. 
In~\se\ref{sc:rate} we compute the leading-order chemical
equilibration rate taking into account the Sommerfeld effect.
A brief summary is presented in \se\ref{se:concl}.

%%%%%%%%%%%%%%%%%%%%%%%%%%%% SECTION %%%%%%%%%%%%%%%%%%%%%%%%%%%%%%%%%%%
%
\section{Non-relativistic QCD in the imaginary-time formalism}
\label{sc:nrqcd}

%%%%%%%%%%%%%%%%%%%%%%%%%%% SUBSECTION %%%%%%%%%%%%%%%%%%%%%%%%%%%%%%%%%
%
\subsection{General formulation}

NRQCD~\cite{nrqcd} describes non-relativistic heavy 
quarks, and gluons and light quarks with momenta much smaller than $ M $. 
Therefore the annihilation of a heavy quark-antiquark pair 
cannot be described 
in terms of the fields of the theory. However, when integrating out 
the scale $ M $, one obtains 4-fermion operators. When this is 
done in real time, their coefficients 
have an imaginary part which corresponds 
to the annihilation process~\cite{bodwin}. 

The imaginary parts of these coefficients arise from a cut, 
or discontinuity, of 4-point functions, viewed  as a function of some
energy variable $ \omega  $, across  the real 
$\omega$-axis. If manipulations are carried out in the complex
$\omega$-plane, as is necessary e.g.\ in thermal field theory, 
then the imaginary
parts of these coefficients 
have to be represented in a way which reflects this analytic
structure. This can be achieved by expressing 
them in a suitable spectral representation. 

The spectral representation of a 2-point function $ \Pi $ can be
written as 
\begin{align} 
 \Pi(\omega,\vc{k}) = 
 \int_{-\infty}^{\infty}
 \! \frac{{\rm d}z}{2\pi}
 \biggl( 
   \frac{1}{\omega - z} - \frac{1}{\omega + z}
 \biggr) \rho(z,\vc{k})
 \;, \la{htl}
\end{align} 
where the spectral density $ \rho $ is a real and odd function of $ z
$.  In this form $ \Pi $ can be evaluated both for real $ \omega $,
which corresponds to real time, and for imaginary $ \omega = i \omega
_ n $, where $ \omega _ n $ is a Matsubara frequency, which
corresponds to imaginary time. By approaching the real $ \omega $-axis
in different ways one obtains different operator orderings
(for instance, 
setting $\omega = \re\omega + i 0^+$ yields a retarded correlator, 
which for $\omega \gg T$ is equivalent to the time-ordered one).
The function $ \Pi ( i \omega_n , \vc k ) $ is
purely real.  Spectral representations are routinely used in finite
temperature perturbation theory for Hard Thermal Loop~\cite{htl1,htl2} 
resummed propagators.

The 4-fermion operators of ref.~\cite{bodwin} are
\ba
 && \hspace*{-2.5cm}
 \delta \mathcal{L}^{ }_M  =  
    \frac{ f_1 ({}^1S^{ }_0)  }{M^2} \,
       \mathcal{O}_1 ({}^1S^{ }_0)
   + \frac{ f_1 ({}^3S^{ }_1)  }{M^2} \,
        \mathcal{O}_1 ({}^3S^{ }_1)
% \nn & & \qquad
   + \frac{ f_8 ({}^1S^{ }_0) }{M^2} \,
         \mathcal{O}_8 ({}^1S^{ }_0)
   + \frac{ f_8 ({}^3S^{ }_1)  }{M^2} \,
   \mathcal{O}_8 ({}^1S^{ }_1)
  \;, \hspace*{0.5cm} \la{four_quark} \\[2mm] 
 \mathcal{O}^{ }_1 ({}^1S^{ }_0) & \equiv & 
  \psi^\dagger \chi \, \chi^\dagger \psi
 \;, \qquad\qquad\;\, 
 \mathcal{O}^{ }_1 ({}^3S^{ }_1) \; \equiv \;  
   \psi^\dagger \vec{\sigma} \chi 
   \cdot \chi^\dagger \vec{\sigma} \psi
 \;, \nn % \la{ops1} \\ 
 \mathcal{O}^{ }_8 ({}^1S^{ }_0) & \equiv & 
   \psi^\dagger T^a \chi \, \chi^\dagger T^a \psi
 \;, \hspace*{0.80cm}
 \mathcal{O}^{ }_8 ({}^3S^{ }_1) \; \equiv \; 
   \psi^\dagger \vec{\sigma} T^a \chi \cdot \chi^\dagger \vec{\sigma} T^a \psi
 \;. \la{ops2}
\ea
Here $\psi,\chi$ are 2-component non-relativistic spinors, 
$\sigma$ are the Pauli matrices, and $T^a$ are generators of 
SU($\Nc$), normalized as $\tr[T^aT^b] = \frac{\delta^{ab}}{2}$. 
The subscripts 1, 8 refer to singlet and octet channels, respectively. 
The absorptive parts of the coefficients read~\cite{bodwin}
\ba
 \im f_1 ({}^1S^{ }_0) & = &  \frac{\CF}{2\Nc}\, 
 \pi \alphas^2 + O(\alphas^3)
 \;, \qquad\;
 \im f_1 ({}^3S^{ }_1) \; = \; O(\alphas^3)
 \;, \nn
 \im f_8 ({}^1S^{ }_0) & = &  \frac{\Nc^2 - 4}{4\Nc}\, \pi \alphas^2 + 
  O(\alphas^3)
 \;, \quad
 \im f_8 ({}^3S^{ }_1) \; = \; \frac{\Nf}{6}\, \pi \alphas^2 + 
  O(\alphas^3)
 \;, \la{imf}
\ea
where $\CF \equiv (\Nc^2-1)/ 2\Nc $.
The corrections of $O(\alphas^3)$ are also known, but not needed here.

The spectral representation of the most general 4-point function 
involves three energy variables instead of a single 
one as in \eq(\ref{htl}). Fortunately,
this complication can be avoided for the operators
of \eq(\ref{ops2}) at 
leading order: 
only the sum of the energies of 
the annihilating particles appears. This is obvious for the $ s $-channel
annihilation. It is also true for the $ t $ and $ u $-channel 
annihilation, because the
virtual heavy quark is far off-shell, and effectively leads to a point-like 
interaction of the annihilating pair and the produced two gluons. 
Consequently, the operators can be represented in 
a form similar to \eq\nr{htl},  
\ba
 \delta {S}_{M}^{(i)}
 & = &  
 \int_ {\mathcal{X}}
 \int_{\mathcal{K}_1,\mathcal{K}_2,\mathcal{K}_3,\mathcal{K}_4}
 e^{i (\mathcal{K}_1+\mathcal{K}_2+\mathcal{K}_3+\mathcal{K}_4)
 \cdot\mathcal{X}} \;
 \psi^*_r(\mathcal{K}_1 ) \,
 \chi^{ }_s(\mathcal{K}_2 ) \,
 \chi^*_t(\mathcal{K}_3 ) \,
 \psi^{ }_u(\mathcal{K}_4 )
 \nn & \times & 
 \int_{-\infty}^{+\infty}
 \! \frac{{\rm d}z}{2\pi}
 \biggl[
   \frac{ \rho^{(i)}_{rstu}(z) }{k^0_1 + k^0_2 - z }
 - 
   \frac{ \rho^{(i)}_{utsr}(z) }{k^0_1 + k^0_2 + z }
 \biggr]
 \;, 
\ea
where $r,s,t,u$ contain both spin and colour indices, 
$i$ enumerates the four cases in \eq\nr{four_quark}, 
$\mathcal{X} \equiv (t,\vc{x})$ and 
$\mathcal{K}_i \equiv (k^0_i,\vc{k}^{ }_i)$. 
% (As is often the case with non-relativistic effective theories,  
% Lorentz invariance is not manifest here, but is 
% still respected by the dynamics.) 
Setting e.g.\
$k_1^0, k_2^0 \to M + i 0^+$, 
$\vc{k}_1, \vc{k}_2 \to \vc{0}$, the absorptive parts can be read off:
\be
 \frac{ \im f_{ }^{(i)} }{M^2} \Leftrightarrow 
 - \fr12 \Bigl[
     \rho^{(i)}_{rstu}(2M) - \rho^{(i)}_{utsr}(-2M)
 \Bigr]
 \;, \la{match}
\ee
where a suitable choice of indices is understood. Subsequently, 
computations can be carried out also in the imaginary-time formalism, 
by including
\ba
 \delta {S}_{E}^{(i)}
 & = & - \;
 \int_X \Tint{ \{ {K}_1,{K}_2,{K}_3,{K}_4 \} }
 e^{i ({K}_1+{K}_2+{K}_3+{K}_4)
 \cdot {X}} \;
 \psi^*_r( {K}_1 ) \,
 \chi^{ }_s( {K}_2 ) \,
 \chi^*_t( {K}_3 ) \,
 \psi^{ }_u( {K}_4 )
 \nn & \times & 
 \int_{-\infty}^{+\infty}
 \! \frac{{\rm d}z}{2\pi}
 \biggl[
   \frac{ \rho^{(i)}_{rstu}(z) }{i k_{n1} + i k_{n2} - z }
 - 
   \frac{ \rho^{(i)}_{utsr}(z) }{i k_{n1} + i k_{n2} + z }
 \biggr]
% \;, 
\la{L_E}
\ea
in  the Euclidean effective action. 
Here 
$X \equiv (\tau,\vc{x})$,
$K \equiv (k_n,\vc{k})$,  
$\int_X \equiv \int_0^{1/T} \!{\rm d}\tau\int_\vc{x}$, 
$\Tinti{\{K\}} \equiv T \sum_{\{ k_n \} }\int_\vc{k}$, 
and $\sum_{\{ k_n \} }$ 
denotes a sum over fermionic Matsubara frequencies. 

%%%%%%%%%%%%%%%%%%%%%%%%%%% SUBSECTION %%%%%%%%%%%%%%%%%%%%%%%%%%%%%%%%%
%
\subsection{Definition of the chemical equilibration rate}
\la{ss:def}

Physically, heavy quark chemical equilibration corresponds to 
the fact that the energy carried by kinetically equilibrated heavy
quarks is not conserved because of annihilation or pair creation; 
the chemical equilibration rate is a 
``transport coefficient'' describing the average non-conservation. 
Concretely, 
the task is to compute the connected correlator~\cite{chemical}
\be
 \Delta(\tau) \equiv 
 \int_{\bf x} 
 \Bigl\langle
    H(\tau,{\bf x}) H(0,{\bf 0}) 
 \Bigr\rangle^{ }_\rmi{c}
 \;, \quad 
 0 < \tau < \frac{1}{T}
 \;, 
\ee
where $H$ denotes the heavy quark Hamiltonian. In terms of 
the fields in \eq\nr{ops2}, the Hamiltonian reads 
$H = M (\psi^\dagger \psi - \chi^\dagger \chi) + O({1} / {M})$.
We expand to first order in the absorptive action, \eq\nr{L_E}, 
which is $1/M^2$-suppressed.
% \footnote{%
% Contact terms may in principle arise when the 4-quark operator is on top of 
% one of the Hamiltonians, but we do not expect these to contribute
% to the cut at the current order.
% } 
After a Fourier transformation, 
$
 \tilde\Delta(\omega_n) = \int_0^{1/T} \! {\rm d}\tau \, 
 e^{i \omega_n\tau} \Delta(\tau)
$, 
and analytic continuation, 
$
 \rho^{ }_\rmii{$\Delta$} (\omega) 
 = \im \tilde\Delta 
 (\omega_n \to -i [\omega + i 0^+])
$, 
a coefficient denoted by $\Omega^{ }_\rmi{chem}$ in ref.~\cite{chemical}
can be extracted as 
\be
 \Omega^{ }_\rmi{chem} = 
 \lim_{\Gamma_\rmii{chem} \,\ll\, \omega \,\ll\, \omega_\rmii{UV}} 
 \omega^2 \bigl[ 1 + 2 \nB{}(\omega) \bigr]
 \rho^{ }_\rmii{$\Delta$} (\omega)  
 \;, \la{Omega_chem}
\ee
where $\Gamma^{ }_\rmi{chem} \sim e^{- M / T} $,  
$\omega_\rmii{UV} \sim  T $, 
and $\nB{}$ denotes the Bose distribution. 
The chemical equilibration rate then follows from 
$
 \Gamma^{ }_\rmi{chem} =
 \Omega^{ }_\rmi{chem} / ( 2 \chi^{ }_{\ff} M^2 ) 
$, 
where $ \chi^{ }_{\ff} $ denotes
the heavy quark-number susceptibility.

%%%%%%%%%%%%%%%%%%%%%%%%%%% SUBSECTION %%%%%%%%%%%%%%%%%%%%%%%%%%%%%%%%%
%
\subsection{Perturbative evaluation of the chemical equilibration rate}

%%%%%%%%%%%%%%%%%%%%%%%%% FIGURE %%%%%%%%%%%%%%%%%%%%%%%%%%%%%%%%%%%%%%%%%
%
\begin{figure}[t]
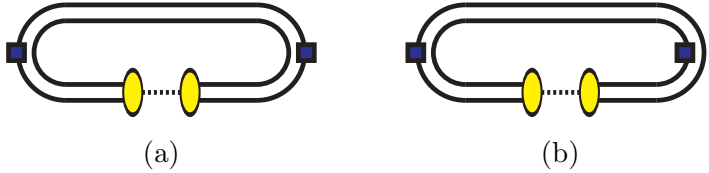


\hspace*{2.2cm}%
\begin{minipage}[c]{11.2cm}
\begin{eqnarray*}
&& 
 \hspace*{-1cm}
 \NRa \qquad\; 
 \NRb \qquad\; 
\\[4mm] 
&& 
 \hspace*{1.4cm}
 \mbox{(a)} \hspace*{4.8cm}
 \mbox{(b)} 
\end{eqnarray*}
\end{minipage}

\caption[a]{\small 
Feynman diagrams for the computation of the heavy quark
chemical equilibration rate within NRQCD. Two ovals connected by 
a dotted line represent the absorptive part of a 4-fermion operator; 
filled squares represent the heavy quark Hamiltonian; and solid 
lines represent heavy quark propagators. The heavy quarks propagate along
the imaginary time direction.}
\la{fig:graphs}
\end{figure}
%
%%%%%%%%%%%%%%%%%%%%%%%%%%%%%%%%%%%%%%%%%%%%%%%%%%%%%%%%%%%%%%%%%%%%%%%%%%

Whereas the formulation of \se\ref{ss:def} 
is in principle non-perturbative (apart from the 
fact that the matching coefficients in \eq\nr{imf}, reflecting ultraviolet
dynamics at the energy scale $\sim M$, 
need to be computed perturbatively), we now 
expand in the gauge coupling as well. The free heavy quark 
propagators have the forms 
\be
 \bigl\langle \psi^{}_r(K_1) \psi^*_s(K_2) \bigr\rangle^{ }_0
 \; = \;  \frac{\delta^{ }_{rs}\, \deltabar(K_1+K_2)}{i k^{ }_{n1} + E^{ }_{k1}}
 \;, \quad
 \bigl\langle \chi^{}_r(K_1) \chi^*_s(K_2) \bigr\rangle^{ }_0
 \; = \; 
 \frac{\delta^{ }_{rs}\, \deltabar(K_1+K_2)}{i k^{ }_{n1} - E^{ }_{k1}}
 \;,
\ee
where $E_k \equiv M + k^2/2M + \cdots $, 
and $\Tinti{K}\deltabar(K) \equiv 1$. 
Feynman diagrams are illustrated in \fig\ref{fig:graphs}.

Carrying out Wick contractions and Matsubara sums, performing the 
analytic continuation, and taking the cut, we obtain the spectral function
\ba
  \rho^{ }_\rmii{$\Delta$} (\omega)  
 & = & \frac{2M^2}{\omega^2}\sum_i
 \int_{\vc{k}_1,\vc{k}_2}
 \Bigl\{ 
  \bigl[
    \rho^{(i)}_{rssr}(E^{ }_{k_1} + E^{ }_{k_2} + \omega) -
    \rho^{(i)}_{rssr}( - E^{ }_{k_1} - E^{ }_{k_2} - \omega) 
  \bigr]
 \nn & \times & 
  \bigl[
   \nF{}(E^{ }_{k_1}) + \nB{}(E^{ }_{k_1} + E^{ }_{k_2} + \omega) 
  \bigr]
  \bigl[
   \nF{}(E^{ }_{k_2} + \omega) - \nF{}(E^{ }_{k_2}) 
  \bigr] \; - \; (\omega \to -\omega)
 \Bigr\} 
 \;, \hspace*{8mm}
\ea
where $\nF{}$ denotes the Fermi distribution. 
Expanding in a small $\omega$, 
\be
   \nF{}(E^{ }_{k_2} + \omega) - \nF{}(E^{ }_{k_2}) 
  \approx 
  -\frac{\omega}{T}\, \nF{}(E^{ }_{k_2}) 
  \bigl[ 1 - \nF{}(E^{ }_{k_2}) \bigr]
 \;,
\ee
and omitting exponentially small terms, the coefficient  
of \eq\nr{Omega_chem} is readily extracted: 
\be
 \Omega^{ }_\rmi{chem} = -8 M^2 
 \sum_i
 \int_{\vc{k}_1,\vc{k}_2}
 %\Bigl\{ 
  \bigl[
    \rho^{(i)}_{rssr}(E^{ }_{k_1} + E^{ }_{k_2}) -
    \rho^{(i)}_{rssr}( - E^{ }_{k_1} - E^{ }_{k_2}) 
  \bigr] \nF{}(E^{ }_{k_1}) \nF{}(E^{ }_{k_2})
 \;.
\ee
Subsequently we may count the contractions for the operators in \eq\nr{ops2}: 
\ba
 \rho^{ }_{rssr}[ 
 \mathcal{O}^{ }_1 ({}^1S^{ }_0)
 ]
 & \to & 
 2 \Nc  
 \;, \qquad\; 
 \rho^{ }_{rssr}[ 
 \mathcal{O}^{ }_1 ({}^3S^{ }_1)
 ]
 \; \to \; 
 6 \Nc
 \;, \nn
 \rho^{ }_{rssr}[ 
 \mathcal{O}^{ }_8 ({}^1S^{ }_0)
 ]
 & \to & 
 2 \Nc \CF 
 \;, \hspace*{4.2mm}
 \rho^{ }_{rssr}[ 
 \mathcal{O}^{ }_8 ({}^3S^{ }_1)
 ]
 \; \to \; 
 6 \Nc \CF
 \;.
\ea
Identifying the absorptive coefficients from \eq\nr{match}
(noting that $E^{ }_{k_1} + E^{ }_{k_2}\approx 2 M$ because of 
the exponential suppression factors), 
and inserting their values from \eq\nr{imf}, then leads to 
\be 
 \Omega^{ }_\rmi{chem}
 = 
 16 M^2 
 \int_{\vc{k}_1,\vc{k}_2}
 \nF{}(E^{ }_{k_1}) \nF{}(E^{ }_{k_2})
 \; \times \; 
 \frac{ \pi \alphas^2 \Nc \CF}{M^2} 
 \biggl( \!\!\!\!
   \underbrace{\frac{1}{\Nc}}_\rmi{singlet, $\mathcal{O}^{ }_1$} \; + \; 
   \underbrace{\frac{\Nc^2 - 4}{2\Nc} + 
   \Nf}_\rmi{octet, $\mathcal{O}^{ }_8$} \;\;
 \biggr)
  \;. \la{split_up}
\ee
This is the main information needed in the next section. 
(For completeness we note that a division by 
$
 2 \chi^{ }_{\ff} M^2 = 8 \Nc M^2 \int_{\vc{k}_2} \nF{}(E^{ }_{k_2})
$
leads to 
$
 \Gamma^{ }_\rmi{chem}
$ of 
\eq\nr{Gamma_chem} with 
$\bSs  = \bSo = 1$.)

%%%%%%%%%%%%%%%%%%%%%%%%%%%%%%%%% SECTION %%%%%%%%%%%%%%%%%%%%%%%%%%%%%%
%
\section{Sommerfeld effect in the chemical equilibration rate}
\label{sc:rate}

Consider now the annihilation or pair creation of a heavy
quark $ Q $ and antiquark $ \bar{Q} $  with four-momenta 
$\mathcal{K}_1 $ and $ \mathcal{K}_2 $. 
We define $v$ as the velocity of $ Q $ 
in the $ Q\bar Q $ rest frame:
\be 
  v \equiv | \vc v |
 \;, \quad
 \vc v \equiv \frac{ \vc k _ 1 -  \vc k _ 2 }{ 2M } 
 \;.
\ee 
One has to
resum the multiple exchange of gluons with typical momenta
$\mathcal{Q}=(q^0,\vc{q})$, where
\begin{align} 
   q ^ 0 \sim M v ^ 2 
   \;, \quad 
   q  \equiv |\vc{q}| \sim M v 
   \;. \label{q-estimate}
\end{align}  
In heavy-quark kinetic equilibrium \eq(\ref{q-estimate}) corresponds to 
\begin{align} 
   q ^ 0 \sim T 
   \;, \quad 
   q  \sim \sqrt{ M T }  
   \;.
\end{align}  
In particular, $ q $ is parametrically larger than the Debye 
scale which is of order $ g %_ {\rm s }
T $, where $ g %_ {\rm s }
$ is the gauge coupling 
($g\equiv \sqrt{4\pi\alphas}$):
\be
  q \gg g T \;. 
\ee
Therefore the Debye screening and Landau damping of 
the exchanged gluons by the hot plasma can be neglected.\footnote{%
 This is true not only parametrically but also numerically: we have 
 checked that, above threshold and for typical parameter values,
 \eq\nr{Ss} is in excellent agreement 
 with the ratio of resummed and tree-level singlet spectral 
 densities~\cite{Og2}, in which the effects of 
 Debye screening and Landau damping are included. 
 }  

The heavy quarks interact with gluons in the plasma, constantly
changing their colour charge. This could affect the Sommerfeld effect
which depends on the colour charge of the pair.  The scattering with
the plasma is characterized by the thermal width $ \gamma  $, which for heavy
quarks is of order $ \alphas  T $ \cite{pisarski}. 
On the other hand, the virtuality 
$
 \Delta \equiv (\mathcal{K} -\mathcal{Q} ) ^ 2 - M^2 = 
  (k^0 - q^0)^ 2 - (\vc k -\vc q ) ^ 2 - M^2
$ 
of the heavy quark lines is of the same order as the
typical momentum transfer squared, i.e.\  
$
 \Delta \sim  M T 
$. 
Schematically, a thermal width would replace 
\begin{align} 
  (k^0 - q^0)^2 \to 
  (k^0 -q^0 + i \gamma  ) ^ 2 \simeq  (k^0 - q^0)^2 + 2 i k^0 \gamma  
  %\;.
\end{align} 
in the propagator.
Since $ k^0 \gamma  \sim \alphas M T  \ll MT \sim \Delta$, 
the width and correspondingly the colour change due to scattering 
with the heat bath  
are small compared with virtuality, and can be neglected at leading order. 

In ref.~\cite{chemical} it was shown that, 
ignoring the Sommerfeld effect, the leading order 
chemical equilibration rate can be obtained 
from a Boltzmann equation which contains the Born cross section. 
The resummation of the Sommerfeld-enhanced terms modifies the Born 
matrix elements as~\cite{sommerfeld,landau3,fadin-top} 
\begin{align}
   |  { \cal M }  _   {\rm resummed } | ^ 2 
   = S \, | { \cal M } _   {\rm tree } | ^ 2 
   \;, \label{resummed}
\end{align} 
where $ S = S( v ) $ is the so-called Sommerfeld factor. 
When the $ Q \bar Q $ pair is in a colour singlet state 
the Sommerfeld factor is $ S = \Ss $ with
\begin{align} 
  \Ss = \frac{ \Xs } { 1 - e ^{ - \Xs } }
  \;, \quad 
  \Xs   = \CF \, \frac{ g ^ 2 } { 4 v }
  \;, \label{Ss}
\end{align} 
whereas for the octet  $ S = \So $ with
\begin{align} 
  \So = \frac{ \Xo } { e ^{  \Xo } -1 }
  \;, \quad    
  \Xo  = \Bigl( \frac{\Nc}{2} - \CF \Bigr) \frac{ g ^ 2 } { 4 v }
  \;. \label{S8}
\end{align}

%%%%%%%%%%%%%%%%%%%%%%%%%%%%%%%% FIGURE %%%%%%%%%%%%%%%%%%%%%%%%%%%
%
\begin{figure}[t]
  \centerline{
    \epsfxsize=7.5cm
    \epsffile{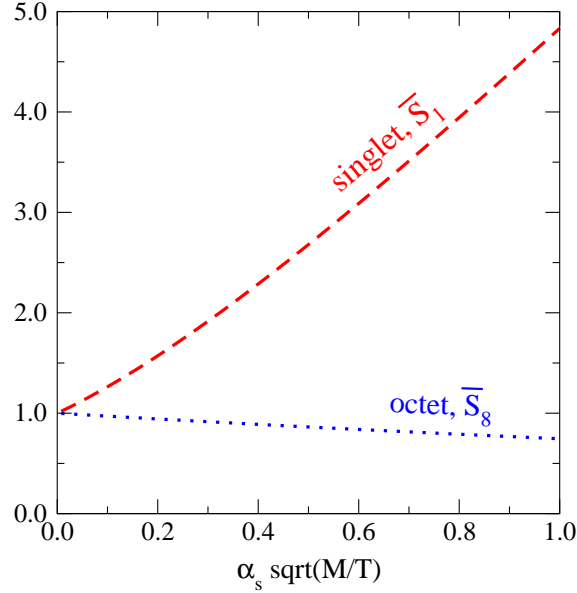}
  } 
  \caption[a]{\small
           The averaged Sommerfeld factors, \eq\nr{barS}, 
           for the singlet and octet contributions.
    \label{fg:factors}
          }
\end{figure}
%%%%%%%%%%%%%%%%%%%%%%%%%%%%%%%%%%%%%%%%%%%%%%%%%%%%%%%%%%%%%%%

At tree level the processes $ gg \leftrightarrow Q \bar Q $ and $
q\bar q \leftrightarrow Q \bar Q $ contribute to the chemical
equilibration rate. The result of ref.~\cite{chemical} can be written as 
\begin{align} 
   \Gamma_\rmi{chem} = \frac{ M e ^{ -M/T } } { 16 \sqrt{ 2 } \pi^ 3 \Nc }
   \int _ 0 ^ {\infty  } \! {\rm d} v \,  v ^ 2 e ^{ - M v ^ 2 /T } 
   \left \{ \fr12 \sum \left | { \cal M } ^ {\rm tree} _ { gg \to Q\bar Q } 
    \right | ^ 2 
   + \Nf \sum \left | { \cal M } ^ {\rm tree}_ { q\bar q \to Q \bar Q }  
   \right | ^ 2 
   \right \} 
   \;, \label{tree}
\end{align} 
where the sums are over all spin and colour degrees of freedom.
In $ q\bar q \leftrightarrow Q \bar Q $ the $ Q \bar Q  $ is in a colour 
octet state, whereas the process $ gg \leftrightarrow Q \bar Q $ has both 
octet and singlet contributions. Denoting by $ r $ the ratio of octet to 
singlet contributions, \eq\nr{split_up} implies
that 
\be 
 r = \frac{\Nc^2 - 4}{2}  =  \fr52 
 \;. \la{r}
\ee 
According to \eq(\ref{resummed}) 
one has to replace in \eq(\ref{tree}) 
\begin{align} 
   \left | { \cal M } ^ {\rm tree} _ { gg \to Q\bar Q }  \right | ^ 2 
   & \to
   \left | { \cal M } ^ {\rm tree} _ { gg \to Q\bar Q }  \right | ^ 2 
   \left ( 
    \frac{ 1 } { 1 + r } \, \Ss 
    + \frac{ r } { 1 + r } \, \So \right ) 
   \;,
   \\
              \left | { \cal M } ^ {\rm tree}_ { q\bar q \to Q \bar Q }  
              \right | ^ 2 
     & \to
              \left | { \cal M } ^ {\rm tree}_ { q\bar q \to Q \bar Q }  
              \right | ^ 2 
              \; \So
   \;.
\end{align} 
The summed tree-level matrix elements in the non-relativistic limit are 
\begin{align} 
   \sum \left | { \cal M } ^ {\rm tree} _ { gg \to Q\bar Q }  \right | ^ 2 
   & = 4 g ^ 4 \CF \Nc ( 4 \CF - \Nc ) 
   \;, \\
   \sum  \left | { \cal M } ^ {\rm tree}_ { q\bar q \to Q \bar Q }  
              \right | ^ 2 
   & = 4 g ^ 4 \CF \Nc
   \;.
\end{align} 
Thus we find
\begin{align} 
   \Gamma_\rmi{chem} =  \frac{ g ^ 4 \CF } { 8 \pi  M ^ 2 }
   \left ( \frac{ M T } { 2 \pi  } \right ) ^{ 3/2 } e ^{ - M/T } 
   \left [ 
      \left ( 2 \CF - \frac{ \Nc } { 2 } \right ) 
      \left ( \frac{ 1 } { 1 + r } \bSs 
              + \frac{ r } { 1 + r } \bSo 
      \right ) 
      + \Nf\, \bSo 
   \right ]        
  \;, \la{Gamma_chem}  
\end{align} 
with the thermally averaged Sommerfeld factors 
\begin{align} 
  \bar S _ \alpha  \equiv \frac{ 4 } { \sqrt{  \pi  } } 
   \left ( \frac{ M } { T } \right ) ^{ 3/2 } 
   \int _ 0 ^ {\infty  } \! {\rm d} v \, v ^ 2 
   e ^{ - M v ^ 2 /T } \, S _ \alpha 
  \;, \quad \alpha \in \{ 1,8 \}
  \;. \la{barS}
\end{align}

After a rescaling of $v$, the Sommerfeld factors
of \eq\nr{barS} are seen to be functions
of $g^2 \sqrt{M/T}$ only. A numerical evaluation is shown in 
\fig\ref{fg:factors}. Analytically, 
 for $ T \ll \alphas^2 M $ we get  
\begin{align} 
   \bSs \approx \frac{g ^ 2 \CF}{2}\sqrt{ \frac{ M } { \pi  T } } 
   \;,
\end{align} 
whereas 
$ \bSo $ is exponentially small (although decreasing 
only slowly in \fig\ref{fg:factors}). 
For $ T \gg \alphas^2 M $, on the other hand, 
\begin{align} 
  \bSs \approx 1 + \frac{ g ^ 2 \CF}{4} \sqrt{ \frac{ M } { \pi  T } } 
   \;, \quad 
  \bSo \approx 1 - \frac{g ^ 2 (\Nc - 2 \CF)}{8}
   \sqrt{ \frac{ M } { \pi  T } } 
   \;.
\end{align} 

As an example, if we take $ \alphas \simeq 0.3 $, 
$ M \simeq 1.5 $ GeV, and $T \simeq 300$~MeV, then 
$\bSs \simeq 3.4$, $\bSo \simeq 0.8$. For $\Nf = 3$, this 
implies that the square brackets in \eq\nr{Gamma_chem} evaluate
to 4.28 rather than the naive 4.17. In other words, the substantial
Sommerfeld enhancement of the singlet channel is all but compensated for 
by the fact that most channels, in particular all associated with 
light quarks, are octets, and for octets there is a mild suppression.  

%%%%%%%%%%%%%%%%%%%%%%%%%%%%%%%%% SECTION %%%%%%%%%%%%%%%%%%%%%%%%%%%%%%%%%%
%
\section{Summary}
\la{se:concl}

In a heavy ion collision, the heavy quark chemical equilibration 
rate parametrizes the rate at which heavy quarks and antiquarks, 
produced in overabundance in an initial hard process, annihilate
during the thermal stage of the fireball evolution. It can be viewed as 
a fundamental property of thermal QCD, whose systematic 
understanding may have 
interesting theoretical relations to cosmology, given 
that similar (co-)annihilation phenomena 
lie e.g.\ at the heart of computations determining the 
dark matter relic abundance (in some scenarios). 

On general grounds, the perturbative expansion for the chemical
equilibration rate has the same functional form as the 
cross section shown in \eq\nr{sketch} with $v\sim \sqrt{T/M}$. In
this paper, we have resummed the terms of 
$O(\alphas ^ n  /v ^ n)$, describing the
Sommerfeld effect, to all orders. The result has the form shown in
\eq\nr{Gamma_chem}, with numerical factors plotted in
\fig\ref{fg:factors}. Due to a fortuitous cancellation between a
strongly enhanced but mildly weighted singlet contribution, and a
mildly suppressed but strongly weighted octet contribution, the
numerical results turn out to be largely insensitive to the
resummation.

The cancellation is peculiar to $\Nc = 3$. For instance, 
for the fundamental representation of SU(2), possibly relevant 
for dark matter (co-)annihilation at temperatures above the 
electroweak scale, the repulsive 
non-singlet contribution is absent (cf.\ \eq\nr{r}). 
There is only an attractive channel also for oppositely charged particles 
in U(1), and indeed the Sommerfeld effect is likely to play 
an important role in chemical equilibration in hot QED plasmas 
(see e.g.\ ref.~\cite{qed} for a general discussion of the problem).

Even though the  $O(\alphas/v)$ contribution
in \eq\nr{sketch} is insignificant in 
practice for $\Gamma_\rmi{chem}$, 
the functions $O(\alphas \ln v)$
and $O(\alphas)$ might well be large. Therefore 
their determination, as well as 
a fully non-perturbative study of the chemical equilibration rate remain, 
in our opinion, well-motivated challenges. 

%%%%%%%%%%%%%%%%%%%%%%%%%%%%%%%%% SECTION %%%%%%%%%%%%%%%%%%%%%%%%%%%%%%%%%%
%
\section*{Acknowledgements}

We thank M.~Beneke, 
X.~Garcia i Tormo, and M.~Garny for helpful discussions. 
The work of D.B.\ was supported in part by the DFG through
the Graduate School GRK 881, and the work of M.L.\ 
by the Swiss National Science Foundation
(SNF) under grant 200021-140234.
D.B.\ thanks ITP/AEC Bern for kind hospitality and support.

%%%%%%%%%%%%%%%%%%%%%%%%%%%%%%% BIBLIOGRAPHY %%%%%%%%%%%%%%%%%%%%%%%%%%%%%%
%

\end{document}